\documentclass[vecphys]{svmult}

\usepackage{makeidx}         
\usepackage{graphicx}        
\usepackage{multicol}        
\usepackage[bottom]{footmisc}

\makeindex             

\begin{document}
\newcommand{\avg}[1]{\langle{#1}\rangle}
\newcommand{\Avg}[1]{\left\langle{#1}\right\rangle}
\def\be{\begin{equation}}
\def\ee{\end{equation}}
\def\bc{\begin{center}} 
\def\ec{\end{center}}
\def\bea{\begin{eqnarray}}
\def\eea{\end{eqnarray}}
\def\bwt{\begin{widetext}}
\def\ewt{\end{widetext}}
\def\ra{\rightarrow}
\def\ba{\backslash}
\title*{Viable flux distribution in metabolic networks}
\author{Ginestra Bianconi\inst{1}\and
Riccardo Zecchina\inst{2}}
\institute{The Abdus Salam International Center for Theoretical
  Physics, Trieste, Italy
\texttt{gbiancon@ictp.it}
\and The Abdus Salam International Center for Theoretical Physics, Trieste,Italy
\texttt{zecchina@ictp.it}}
%
%

\maketitle

\begin{abstract}
The metabolic networks are very well characterized  for a
large set of organisms, a unique case in within thelarge-scale  biological
networks.
For this reason they provide a a very interesting framework for the
construction of analytically tractable statistical mechanics models.
 In this paper we introduce a solvable model for the distribution of
fluxes in the  metabolic network. 
We show that the effect of the topology on the
distribution of fluxes is to allow for large fluctuations of their
values, a fact that should have implications on the robustness of the system. 
\end{abstract}
\section{Introduction}
\label{sec:intro}

Dynamical models on networks have attracted  a large interest
because of the non-trivial  effects of network structure
\cite{RMP,Doro,Newman,Internet}  on the dynamics  defined on them \cite{Critical}.
Important examples of the dynamics on networks with
relevant applications are the  Ising
model \cite{Doro_Ising,Diluito,annealed}, 
the spreading of a disease \cite{Vespi_epi} and the synchronization models \cite{Motter,Moreno}.
In this paper introduce a solvable model for the distribution of
fluxes in the  metabolic network.
 While motivations come  from the study of the
metabolic network, the problem is quite
general and can be applied to supply networks and to many other linear
problems \cite{Linear} of constraint satisfaction on continuous variables on a
network. 

Metabolic networks  describe the stoichiometric relations between
 substrates in biochemical reactions inside the cell. They have been
 mapped \cite{BIGG} for a large number of organisms in the three different domains
 of life (archea, bacteria and eukaryotes). 
 They  provide the biomass needed for  cell duplication, and the rate of biomass production (growth rate) can be identified with a fitness of the cell.
The structure of the metabolic  network can be represented as  a factor graph
with nodes that are chemical reactions and function nodes that are
chemical metabolites.
The  projection of the network on the metabolites  has a power-law degree distribution and a hierarchical structure \cite{metabolic,modularity,Tanaka}.
To each factor node, which describes a chemical
reactions, it is  associated an enzyme which itself is produced by a regulated gene network. Important aspect of the functioning of these very complex systems include dynamical considerations. 
Flux-balance-analysis \cite{fba,Palsson,Palsson2} make a major
semplification in the problem. In fact it considers  only the steady
state of the dynamics and includes all the dynamical terms inside the
definition of the flux of a reaction. For this reason it  was  able to
predict with sufficient accuracy the fluxes of the reactions in the
graph for a given  environmentand it consitute a real break-through in
the field.
Special interest has been addressed to the perturbation of the distribution of the fluxes after knockout of a gene or in different environments \cite{MOMA,ROOM}.
The problem of identifying the flux distribution in {\it Escherichia
  coli} was studied experimentally \cite{exp} and by means of 
Flux-Balance-Analysis \cite{Almaas} . A fat tail in their distribution with
different power-law exponents $\alpha<2$ was found.

Metabolic networks provide a very interesting framework for the construction of
analytically tractable models  using tools of statistical mechanics of
disordered systems. 
In this paper  we  will   discuss the impact of the network structure (degree
distributions) on the steady state distribution distribution of the fluxes. 
We shall  consider  random networks with
the same degree distribution as the  real ones 
i.e.  networks in the the
hidden-variable ensemble \cite{hv1,hv2,hv3} with same expected degree
distribution as the metabolic factor graphs.
Formally the problem is resolved with replica calculations on diluted
networks \cite{Diluito} extended to the case of   continuous variables.  Due the simplicity of the
Hamiltonian  the problem is solved with an expansion of the order
parameter in terms of Gaussians.
The problem shares some similarity with other problems in statistical
mechanics of disordered systems \cite{Localized,Guil}.
In a   recent paper \cite{DeMartino} a similar problem  was
considered
in the framework of a different  model where the steady state of the 
fluxes is not a priori considered and the positive fluxes  don't have any upper limit.

\section{The model}
\label{sec:model}                   
The metabolic network has a bow tie structure \cite{Tanaka} : 
the metabolites are
divided into: {\it(i)} input metabolites which are provided by the environment,
{\it (ii)} the output metabolites which provide the biomass and {\it
  (iii)} the intermediate
metabolites.
The stochiometric matrix is given by $((\xi_{\mu,i}))$ where
$\mu=1,\dots,M$ indicates the   metabolite and $i=1,\dots,N$
the   reaction and the sign of $\xi_{\mu,i}$ indicates if the
metabolite $\mu$ is an input or output metabolite of the reaction $i$.
As in the Flux-Balance-Analysis method we assume that each intermediate
metabolite has a concentration $c_{\mu}$  which is consumed/produced
by a  reaction $i$  at a rate $f_i$.  At steady state, we have
\be
\frac{d c_{\mu}}{dt}=\sum_i \xi_{\mu,i} f_i=a^{\mu}=0
\label{ss.eq}
\ee
where $f_i$ is the flux of the metabolic reaction $i$.
For the metabolites present in the environment and the metabolites
giving rise to the biomass production we can fix  the 
incoming flux given by
\be
\frac{d c_{\mu}}{dt}=\sum_i \xi_{\mu,i} f_i=a^{\mu}_{in/out}.
\label{ss2.eq}
\ee
The fluxes in Eqs.(\ref{ss.eq}-\ref{ss2.eq}) can vary inside a fixed
volume $\Omega$.
We assume for simplicity that this volume is an hypercube
$\Omega=[0,2L]^N$.
Changing the variables $f_i$ in the variables $s_i=f_i-L$ and  the equations that the fluxes $s_i$ must satisfy are given by 
\be
\sum_{i=1}^N \xi_{\mu,i} s_i=g_{\mu}\  \mbox{for}\  \mu=1,\dots M.
\label{uno}
\ee
where $g_{\mu}=a^{\mu}-L\sum_{i} \xi_{\mu,i}$.
The  volume of solutions $V$, given the constraints $(\ref{uno})$, 
 is proportional to the quantity 
\be
\tilde{V}=\int_0^{L} dL'\int \prod_{i=1}^{N} ds_i \delta(\sum_i \xi_{\mu,i} s_i-g^{\mu})
 \delta(\sum_j q_j s_j^{2} -\avg{q_i}NL'^2).
\ee
where we have used   the heterogeneous
spherical constraints
\be
\frac{1}{\avg{q_i}N}\sum_i q_is_i^2=L'^2 
\label{sph}
\ee
and integrated over $L'$ in the interval $[0,L]$ in order to allow
analytical treatment of the problem.

\section{Replica method }
\label{sec:replica}

We assume that the support of our stochiometric matrix is a random 
network with given degree distribution, i.e. a realization of the
random hidden-variable model \cite{hv1,hv2,hv3}.
In particular we fix the expected degree distribution of the nodes of
the factor graphs to be  $q_{i}$ for the reaction node $i=1,\dots N$
and  $q_{\mu}$ for the metabolite nodes $\mu=1,\dots, M$ and we assume
that the matrix elements $\xi_{\mu,i}$ are distributed following 
\be
P(\xi_{\mu,i})=\frac{q_{i}q_{\mu}}{2\avg{q_i}N}
\left[\delta(\xi_{\mu,i}-1)+\delta(\xi_{\mu,i}+1)\right]+\left(1-\frac{q_{i}q_{\mu}}{\avg{q_i}N}\right)
\delta(\xi_{\mu,i}),
\label{pxi}
\ee
where $\delta()$ indicates the Kronecker delta.
Note that in   $(\ref{pxi})$ we have assumed that the elements of the 
stochiometric matrix  have values  $0,\pm1$  with a random sign and
that  the variables $q_{i}$ $q_{\mu}$ are nothing else than the
hidden-variables  associated with
metabolite $\mu$ of the reaction $i$ of the hidden-variable  network
ensemble
\cite{hv1,hv2,hv3}.

In order to evaluate the steady state distribution of the fluxes  in a typical network realization we 
replicate the realizations of the $s_i^a$ and we compute
$\avg{\log({V})}$ over the different network realizations.
To calculate this average we use  the  replica trick $S=\avg{\log({Z})}=\lim_{n\rightarrow
  0} \frac{\avg{\tilde{V}^n}-1}{n}$.
The averaged unormalized volume of solutions  $<\tilde{V}^n>$ can be
expressed as 
\bea
<\tilde{V}^n>=\int_0^{L} dL' \int \prod_a d{\omega^a}\int \prod_{a,i}
d s_{i,a} \int \prod_{a,\mu} d\lambda_{\mu,a} \exp\left[-i
  g_{\mu}\sum_a \lambda_{\mu,a}\right]\nonumber \\
 \exp\left[-\sum_{i,\mu}\frac{q_i
    q_{\mu}}{\avg{q_i} N}(1-\cos  {\vec{\lambda}}_{\mu} \cdotp
  \vec{s}_i)+i\sum_a \omega^a \left(\sum_j q_js_{j,a}^2-L'^2\avg{q_i}N\right)\right].
\eea
Using the techniques coming from the field of diluted systems, we
introduce the order parameters \cite{Order,Diluito}

\bea
c(\vec{\lambda})=\frac{1}{\avg{q_i} N} \sum_{\mu}q^{\mu} \prod_a \delta(\lambda_{\mu}^a-\lambda^a) \nonumber\\
c(\vec{s})= \frac{1}{\avg{q_i}N} \sum_{i}q^i \prod_a \delta(s_{i}^a-s^a)
\eea
getting  for the volume

\bea
\Avg{\tilde{V}^n}=
\int {\cal D}c(\vec{\lambda}) \int {\cal{D}}\hat{c}(\vec{\lambda})
\int {\cal D}c(\vec{s})\int  {\cal{D}}\hat{c}(\vec{s})
\exp[nN\Sigma(\hat{c}(\vec{\lambda}),c(\vec{\lambda}),\hat{c}({\vec{s}}),c({\vec{s})})]\nonumber 
\eea

with
\bea
n\Sigma &=&\int  d\vec{\lambda} i \hat{c}(\vec{\lambda}) c(\vec{\lambda})+\int  d\vec{s} i \hat{c}({\vec{s}}){c}({\vec{s}})- \int d\vec{\lambda}\int  d\vec{s} c(\vec{\lambda}) c({\vec{s}})(1-\cos(\vec{\lambda} \cdotp {\vec{ s}}))+\nonumber \\
&+&\frac{1}{\avg{q_i}N} \sum_{\mu} \log \int d\vec{\lambda} \exp[-i
g_{\mu}\sum_a \lambda_a
-iq_{\mu}\hat{c}(\vec{\lambda})]-i\sum_a\omega^a L'^2\nonumber \\ &+&\frac{1}{\avg{q_i}N}\sum_i\log\int  d \vec{s} \exp[-iq_i \hat{c}({\vec{s}})+i\sum_a q_i\omega_a s_a^2].\nonumber 
\eea

The saddle point equations for evaluating $\Sigma$ are given by
\bea
i\hat{c}(\vec{\lambda})&=& \int  d\vec{s} c({\vec {s}}) (1-\cos({\vec{\lambda} \cdotp \vec{s})})\nonumber \\
i\hat{c}({\vec{ s}})&=& \int  d\vec{\lambda} c(\vec{\lambda}) (1-\cos({\vec{ \lambda}\cdotp \vec{s}}))\nonumber \\
c(\vec{\lambda})&=&\frac{1}{\avg{q_i}N} \sum_{\mu}
q_{\mu}\frac{\exp\left[-i g^{\mu}\sum_a \lambda_a-i q_{\mu}\hat{c}(\vec{\lambda})\right]}{\int \prod_a d\lambda'_a\exp\left[-ig_{\mu}\sum_a \lambda'_a-i q^{\mu}\hat{c}(\vec{\lambda}')\right ]}\nonumber \\
c(\vec{s})&=& \frac{1}{\avg{q_i}N}\sum_i q_i\frac{\exp\left[-i q_i\hat{c}(\vec{s}) +i\sum_a \omega_a s_a^2\right]}{\int d\vec{s}' \exp\left[-i q_i\hat{c}(\vec {s}') +i\sum_a \omega_a {( s')}_a^2\right]}\nonumber \\
L^2&=&\frac{1}{\avg{q_i}N} \sum_i q_i\frac{\int d\vec{s} s_a^2
  \exp[-iq_i \hat{c}({\vec{s}})+i\sum_a q_i\omega^a s_a^2] }{\int d
  \vec{s}' \exp[-iq_i \hat{c}({\vec{s}})+i\sum_a q_i\omega_a s_a^2]}.
\label{sp.eq}
\eea
We assume that the solution of the saddle point equation is replica
symmetric, i.e. the distribution of the variables $z^a=\lambda^a,s^a$ 
 conditioned
to a vector field ${\vec{x}}$ are identically equal distributed,
\be
c(\vec{z})=\int d\vec{x} P(\vec{x})\prod_{a=1}^n \phi(z^a|\vec{x})
\ee
where $\phi(z|\vec{x})$ are distribution functions of $z$ and
$P({\vec{x}})$ is a probability distribution  of the vector field
$\vec{x}$.
For the  function $\phi(z|\vec{x})$ the exponential form is usually assumed 
in Ising models.
In our continuous variable case  for our quadratic problem, we assume
instead that  $\phi(z|\vec{x})$  has a Gaussian form.
This assumption could be in general considered as an approximate solution
of the equations $(\ref{sp.eq})$.
Explicitly we   assume that the functions $c(\vec{\lambda})$ and
$c(\vec{s})$ can be expressed as  the following,
\bea
c(\vec{\lambda})&=&\int dm_{\lambda} dh_{\lambda} P(h_{\lambda},m_{\lambda})\prod_a \exp\left [-\frac{1}{2}h_{\lambda} \lambda_a^2+\frac{1}{2} \frac{m_{\lambda}^2}{h_{\lambda}} \right] \cos[m_{\lambda}\lambda_a]\sqrt{\frac{h_{\lambda}}{2\pi}}\nonumber \\
c({\vec{ s}})&=&\int dm_{{ s}} dh_{ {s}} P(h_{ s},m_{ s})\prod_a \exp\left[-\frac{1}{2}h_{s} s_a^2-\frac{1}{2}\frac{m_s^2}{h_s}\right] \cosh[m_s s^a] \sqrt{\frac{h_s}{2 \pi}}\nonumber \\
\omega_a&=&i\omega
\label{gaus1.eq}
\eea
from which we derive for $\hat{c}(\vec{s})$ and
$\hat{c}(\vec{\lambda})$
\bea
\hat{c}(\vec{s})&=&-i\left(1-\int dm_{\lambda} dh_{\lambda} P(h_{\lambda},m_{\lambda})\prod_a \exp\left [- \frac{1}{2 h_{\lambda}}s_a^2\right] \cosh[m_{\lambda} s_a]/h_{\lambda}\right) \nonumber \\
\hat{c}(\vec{\lambda})&=&-i\left(1-\int dm_{ s} dh_{ s} P(h_{s},m_{s})
  \prod_a \exp\left[-\frac{1}{2 h_{s}} \lambda_a^2\right] \cos[m_{s}
  \lambda^a/h_{s}]\right).\label{gaus2.eq}
\eea
The saddle point equations $(\ref{sp.eq})$, taking into account the
expression for the order parameters $(\ref{gaus1.eq})(\ref{gaus2.eq})$
 closes and can be written  as recursive equation for $P(h_{\lambda},m_{\lambda})$
and $P(h_s,m_s)$, i.e.

\bea
P(h_{\lambda},m_{\lambda})&=&\frac{1}{\avg{q_i}N}\sum_{\mu}q_{\mu}\sum_k
e^{-q_{\mu}} q_{\mu}^k\frac{1}{k!}\int_{\dots}\int \prod_i dh_{s}^i
dm_{s}^i\prod_i P (h^i_{s},m^i_{s})\nonumber \\
& &\delta\left(h_{\lambda}-\sum_{i=1}^k \frac{1}{h_{s}^i}\right) \frac{1}{2^k}\sum_{\{n_i\}} \delta\left(m_{\lambda}-\sum_i (-1)^{n_i} \frac{m_{s}^i}{h_s^i}-g_{\mu}\right) \nonumber \\
P(h_s,m_s)&=&\frac{1}{\avg{q_i}N}\sum_i q_i e^{-q_i}\sum_k
q_i^k\frac{1}{k!}\int_{\dots} \int  \prod_i dh_{\lambda}^i dm_{\lambda}^i
\prod_i P(h^i_{\lambda},m^i_{\lambda})\nonumber \\
& & \delta\left(h_s-\sum_{i=1}^k
  \frac{1}{h_{\lambda}^i}-2  \omega q_i\right) \frac{1}{2^k}\sum_{\{n_i\}} \delta\left (m_s-\sum_i (-1)^{n_i} \frac{m_{\lambda}^i}{h_{\lambda}^i}\right) \label{recursive}\\
L^2&=&\frac{1}{\avg{q^i} N}\sum_iq^i e^{-q^i}\sum_k
\frac{q_i^k}{k!}\sum_{s_i}\frac{1}{2^k}\int_{\dots}\int\prod_i
dh_{\lambda}^i dm_{\lambda}^i
\prod_iP(h_{\lambda}^i,m_{\lambda}^i)\nonumber\\ & &\frac{(H+M^2)} \delta\left(H-\sum_{i=1}^k \frac{1}{h_{\lambda}^i}-2  \omega q_i\right)
\frac{1}{2^k}\sum_{\{n_i\}} \delta\left (M-\sum_i (-1)^{n_i}
  \frac{m_{\lambda}^i}{h_{\lambda}^i}\right).\nonumber 
\eea

Finally $S$ can be calculated at the saddle
point as

\bea
S&=&-\int  dh_{s} dm_{s}  dh_{\lambda}
dm_{\lambda}P
(h_{s},m_{s})P(h_{\lambda},m_{\lambda})\left[-\frac{(m_s/h_s)^2}{2(h_{\lambda}+1/h_s)}+\frac{(m_{\lambda}/h_{\lambda})^2}{h_s+{1}/{h_{\lambda}}}+\right.\nonumber\\
&&\left.+\log
  \cosh \left(\frac{m_s m_{\lambda} }{h_s
      h_{\lambda}+1}\right)\right]+\label{S} \\ & &
+ \frac{1}{\avg{q^i}N}\sum_{\mu}\sum_k e^{-q_{\mu}}
q_{\mu}^k\frac{1}{k!}\int \prod_i dh_{s}^i dm_{s}^iP
(h^i_{s},m^i_{s}) \frac{1}{2^{k+2}} \left[
   \frac{ g_{\mu}^2+\sum_i  (m_{s,i}/h_{s,i})^2}{\sum_j \frac{1}{h_s^j}}\right]\nonumber \\
& & +\frac{1}{\avg{q_i}N}\sum_i \sum_k e^{-q_i}\ q_i^k\frac{1}{k!}\int
\prod_i dh_{\lambda}^i dm_{\lambda}^i
P(h^i_{\lambda},m^i_{\lambda})\frac{1}{2^{k+1}}\left[\frac{\sum_i (m_{\lambda,i}/h_{\lambda,i})^2}{\sum_i \frac{1}{h_{\lambda,i}}+2 \omega q_i}\right]\nonumber
\label{entropy}
\eea

\section{Population Dynamics}
\label{sec:popul}

We solved the equations $(\ref{recursive})$ by a population-dynamical algorithm.
We represent the effective field distributions $(h_s,m_s)$
$(h_{\lambda},m_{\lambda})$ by a large population of $P\gg1$ fields. Running the algorithm the population is
first initialized randomly and then equations $(\ref{recursive})$  are
used to iteratively replace the fields  inside the population until
convergence is reached.
Instead of fixing $\omega$ we introduce a further variable $\Lambda$
fixing the value of the average flux in the network.
 The action of the algorithm is summarized in the following pseudo
 code
\\
\\
{\bf algorithm} PopDyn($\{h_s^1,m_s^1,h_s^2,m_s^2,\dots,h_s^P,
m_s^P\}$;\\\
$\{h_{\lambda}^1,m_{\lambda}^1,h_{\lambda}^2,m_{\lambda}^2,\dots,h_{\lambda}^P,
m_{\lambda}^P\},\omega$)
\\
{\bf begin}\\
{\bf do }\\
\begin{itemize}
\item choose a metabolite $i_0$ with probability $q_{i} P(q_i)$;\\
\item draw {\it d} from a Poisson distribution ($e^{-q_i}q_i^k/k!$)\\
\item select $d$ indexes $i_1,\dots i_d\in\{1,\dots M\}$\\
\item draw a $d$-dimensional  vector $\vec{n}=\{n_i\}$ of random numbers $n_i=0,1$\\
\bea
h_s^{i_0}:&=&\sum_{l=1}^d\frac{1}{h_{\lambda}^{i_l}}+2\omega q_i;\nonumber \\
m_s^{i_0}:&=&\sum_{l=1}^d (-1)^{n_i}\frac{m_{\lambda}^{i_l}}{h_{\lambda}^{i_l}};\nonumber\\
L_2:&=&\left(1-\frac{1}{\avg{q^i}N}\right)L_2+\frac{1}{\avg{q_i}N}
  \frac{h_s^{i_0}+(m_s^{i_0})^2}{(h_s^{i0})^2};\nonumber\\
\omega:&=&\frac{L_2}{\Lambda^2}
\eea
\item choose a random reaction  $\mu_0$ with probability $q_{\mu} P(q_{\mu})$\\
\item draw {\it d} form a Poisson distribution ($e^{-q_{\mu}}q_{\mu}^k/k!$)\\
\item select $d$ indexes $i_1,\dots i_d\in\{1,\dots M\}$\\
\item draw a $d$-dimensional vector $\vec{n}=\{n_i\}$ of random numbers $n_i=0,1$\\
\bea
h_{\lambda}^{\mu_0}:&=&\sum_{l=1}^d\frac{1}{h_{s}^{i_l}}+2\omega;\nonumber \\
m_{\lambda}^{\mu_0}:&=&\sum_{l=1}^d (-1)^{n_i}\frac{m_{s}^{i_l}}{h_s^{i_l}}+g^{\mu_0}
\eea
\end{itemize}
{\bf while} (not converged)\\
{\bf return}\\
{\bf end}
\\
\\

We run the population dynamics algorithm and 
we measure the distribution of the average fluxes $m_s/h_s$, the
distribution of the fields $h_s$  for
different values of $\Lambda$. We consider as the underline network a
network with   the real degree distribution of the
metabolic factor graph of Saccharomyces cerevisiae and on a network with
the same number of metabolites and reactions as the real Saccahromyces
cerevisiae network but with a fixed connectivity for each metabolite and reaction node.
We consider a   population of  $P=3000$ pair of fields $(h_s,m_s)$.
A random fraction of $5\%$ of the nodes is chosen as an
input/output metabolite.
The values of $g^{\mu}$ are chosen randomly depending if the
metabolite $\mu$ is an intermediate metabolite or an input/output metabolite.

For the input/output metabolites  we assume that $g^{\mu}$ is 
 a random number uniformly distributed in the interval
$[-100\Lambda ,100\Lambda]$  mimicking high rate of production/consumption. For
intermediate metabolites  we choose $g^{\mu}$ with a uniform distribution in the interval
$[-\Lambda, \Lambda]$.

\begin{figure}
\centering
\includegraphics[height=8cm]{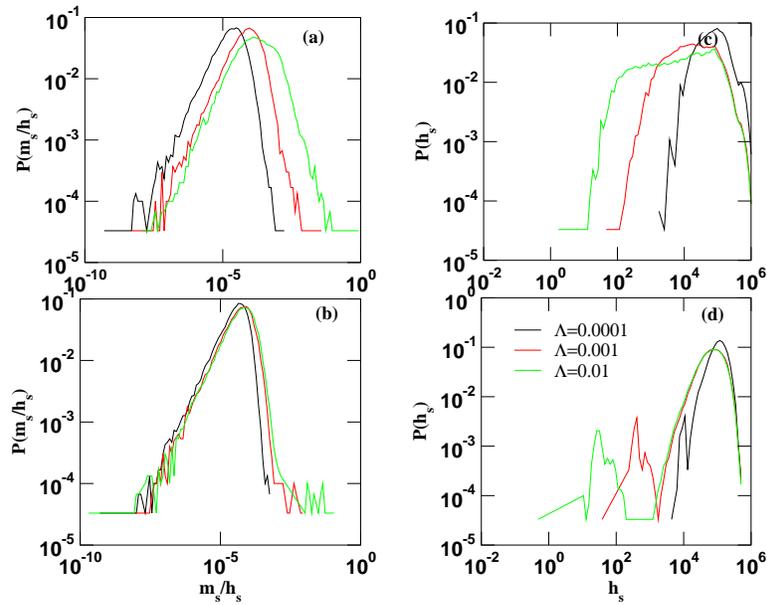}
%
%
\caption{Distribution of the fields $m_s/h_s$ and of the fields $h_s$
  for a the real degree distribution of the metabolic network of
  Saccharomyces cerevisiae (graphs (a) and (c))and for a graph with
  the same number of metabolites and reactions and the same number of nodes that the real metabolic network of Saccharomyces cerevisiae but with  two delta peaks for the degree distribution (graphs (b) and (d)). }
\label{replica.fig}       
\end{figure}

The distribution of $m_s/h_s$   as a function of $\Lambda$ are plotted in figure 
$\ref{replica.fig}$(a) for the real metabolic network of Saccharomyces
cerevisiae and in figure $\ref{replica.fig}$(b) 
for the random network with two delta function degree distribution
($P(q_i)=\avg{q_i}$, $P(q_{\mu})=\avg{q_i}N/M$).
 We observe that the  average fluxes
distribution $m_s/h_s$ in Saccharomyces cerevisiae for low $\Lambda$
has a fatter tail for the real degree distribution than for the two
delta peak degree distribution.

On the other side the distribution of $h_s$ is very different in the real and in the random case (see figure $\ref{replica.fig}$(c)-(d)).
 In particular for the real metabolic network degree distribution is
  broader allowing with higher probability for smaller value of $h_s$
  than in  the case of a two delta peak degree distribution. Therefore
  we have shown that the real topology of the networks has  as a
major consequence in allowing larger  fluctuations of the fluxes in the network.

%
\section{Conclusions}
In this paper we have proposed a statistical mechanics
approach for the study of flux-balance-analysis in a particular
ensemble of metabolic networks. We have studied the impact of the
topology of the networks on the distribution of the fluxes.
We observe that the role of the real topological structure is to
allow for larger variation of the  fluxes, a fact that should have
implications for the robustness of the system. In particular we found
that the topology of  real metabolites has
 an impact on the fat tail of the $m_s/h_s$ distribution
and on the small $h$ field  of  the network, 
Further work is under consideration for the implementation of a message-passing algorithm able to predict the fluxes taking into account the full complexity of the real metabolic network.

\section{Acknowledgments}
We acknowledge  E. Almaas, A. De Martino, R. Mulet and  G. Semerjian for stimulating discussions.
%
%
%
%
%

%
%







\printindex
\end{document}